\documentclass[compsoc,conference,a4paper,10pt,times]{IEEEtran}
\IEEEoverridecommandlockouts
\usepackage{cite}
\usepackage{amsmath,amssymb,amsfonts}
\usepackage{algorithmic}
\usepackage{graphicx}
\usepackage{textcomp}
\usepackage{bmpsize}
\usepackage{xcolor}
\usepackage{lipsum}
\usepackage[colorlinks=true,urlcolor=black]{hyperref}
\def\BibTeX{{\rm B\kern-.05em{\sc i\kern-.025em b}\kern-.08em
    T\kern-.1667em\lower.7ex\hbox{E}\kern-.125emX}}

\usepackage{multirow}
\usepackage{subfig}
\begin{document}

\title{GNPassGAN: Improved Generative Adversarial Networks For Trawling Offline Password Guessing}

\author{\IEEEauthorblockN{Fangyi Yu,
Miguel Vargas Martin}
\IEEEauthorblockA{Ontario Tech University\\
Oshawa, Canada\\
\{fangyi.yu, miguel.martin\}@ontariotechu.ca}}

\maketitle

\begin{abstract}
The security of passwords depends on a thorough understanding of the strategies used by attackers. Unfortunately, real-world adversaries use pragmatic guessing tactics like dictionary attacks, which are difficult to simulate in password security research. Dictionary attacks must be carefully configured and modified to represent an actual threat. This approach, however, needs domain-specific knowledge and expertise that are difficult to duplicate. This paper reviews various deep learning-based password guessing approaches that do not require domain knowledge or assumptions about users' password structures and combinations. It also introduces GNPassGAN, a password guessing tool built on generative adversarial networks for trawling offline attacks. In comparison to the state-of-the-art PassGAN model, GNPassGAN is capable of guessing 88.03\% more passwords and generating 31.69\% fewer duplicates.
\end{abstract}


\section{Introduction}
Passwords have dominated authentication systems for decades, despite their security flaws compared to competing techniques such as cognitive authentication and hardware tokens. Their prevalence is primarily due to their incomparable deployability and usability \cite{bonneau2012quest}. However, the security of user-selected passwords continues to be a significant concern. According to research examining susceptible behaviours that affect password crackability \cite{adams1999users}, there are three types of user actions that result in the creation of insecure passwords: (1) Users often use basic terms in passwords and perform simple string transformations to comply with websites' password creation policies \cite{narayanan2005fast}. (2) Password reuse is prevalent. According to S. Pearman et al. \cite{pearman2017let}, 40\% of users reuse their passwords across multiple platforms. (3) Users prefer to use simple-to-remember passwords that include personal information such as their birth date or their pets' name. All of these behaviours expose the user-created passwords to attacks. Additionally, the recent large-scale leakage of passwords on multiple platforms across the world (listed in Table \ref{tab:datasets}) raises the alarm for researchers and stakeholders.  

\begin{table}[t]
  \begin{center}
    \caption{Real passwords obtained from data breaches that are often used in password guessing experiments. The size column indicates the total number of unique passwords in the dataset}
    \label{tab:datasets}
    \begin{tabular}{ccc}
    \hline
      \textbf{Name} & \textbf{Size} & \textbf{Brief Description} \\
    \hline
      Yahoo\ & $4.4\times10^5$ & A web services provider.\\ 
      phpBB &  $3.0\times10^5$ & A software website.\\ 
      RockYou\ & $1.4\times10^7$ & A gaming platform.\\ 
      Myspace\ & $5.5\times10^4$ & A social networking platform.\\ 
      SkullSecurityComp\ & $6.7\times10^6$ & Compilations of password lists.\\ 
      LinkedIn & $1.3\times10^6$ & A social online platform.\\
      \hline
    \end{tabular}
  \end{center}
\end{table} 

As a consequence, it becomes even more critical to aid users in establishing stronger passwords. Due to the fact that password strength is a statistic that reflects a password's resistance to guessing attacks, the first step is to appropriately estimate password strength via password guessing attacks. Password guessing strategies can be categorized as offline and online, or targeted and trawling. Our study concentrates on the offline, trawling situation, as justified in Sections \ref{OO} and \ref{TN}.

\subsection{Offline Attacks and Online Attacks}\label{OO}
Offline attacks occur when attackers get cryptographic hashes of certain users' passwords and attempt to recover them by guessing and testing many passwords. The primary objective is to determine the difficulty of cracking a genuine user's password, or the strength of a user-created password, by producing a list of password guesses and checking for the possibility of the genuine user's password's occurrence. Offline attacks are only considered when the following conditions are met: An attacker gains access to the system and extracts the password file, all while remaining unnoticed. Moreover, the file's salting and hashing must be done appropriately. Otherwise, an offline attack is either ineffective (the attacker may get credentials directly without requiring guesses, or an online approach is more effective), or infeasible \cite{florencio2014administrator}. 

An online attack occurs when an attacker makes password attempts against users using a web interface or an application. This situation is more constrained for attackers since most authentication systems automatically freeze accounts after several unsuccessful attempts. Therefore, attackers must guess users' passwords successfully within the allotted number of tries, which is the primary difficulty of online password guessing. 

According to Flor\^{e}ncio et al. \cite{florencio2014administrator}, $10^6$ is a reasonable upper limit for the number of online guesses a secure password must survive, while the number of offline guesses is difficult to quantify considering the attacker's possible usage of unlimited computers with each calculating hashes thousands of times quicker than the target site's backend server. One of the most significant advantages of deep learning models is their ability to produce a significant amount of passwords that follow the distribution of real passwords. As the number of generated passwords rises, the number of passwords correctly guessed likewise increases. Therefore, our study will concentrate on offline attacks in order to maximize the benefits of deep learning models.

\subsection{Targeted Attacks and Trawling Attacks}\label{TN}
Targeted guessing attacks take place when attackers
try to crack users’ passwords using data containing any information related to the user. These information are classified into two categories according to their degree of confidentiality \cite{wang2016targeted}. The first type is user personally identifiable information, including name, email and gender, and is available online, e.g., through social media.  The second type is user identification credentials, which are partially public (e.g., username) and partially private (e.g., password). Trawling attacks, on the other hand, try to find a user account that matches a known password, so they do not presume the users' identities and are thus more omnipresent.
Be aware that the majority of individual accounts do not deserve concentrated attention. Only significant or vital accounts relating to critical work, finances, or documents demanding a high level of security may be considered for targeted attacks \cite{florencio2014administrator}.

In these regards, this paper primarily focuses on utilizing deep learning to increase passwords guessability under the offline and trawling scenario.
\subsection{Contributions}

The following are the primary contributions of this paper: 
\begin{itemize}
\item We provide an in-depth literature review of the deep learning techniques used for trawling offline password guessing. 
\item We introduce GNPassGAN, which outperforms standalone password guessing methods in the literature on both one-site (models trained and tested on subsets of the same dataset) and cross-site (models trained and tested on various datasets) scenarios. It can serve as (a) a new state-of-the-art benchmark for academics interested in the password guessing area, and can further be utilized to (b) develop password strength meters that encourage users to choose stronger passwords, and (c) to produce honeywords that detect password breaches. \cite{juels2013honeywords}. 
\item  We highlight issues with password settings when deep learning algorithms are used to guess passwords. More specifically, previous research mainly focused on guessing passwords less than or equal to 8 characters, which are considered short passwords and would be rejected by websites with a minimum length password policy. This enables future research to focus only on password settings that adhere to password creation policies.
\item To encourage more study on password guessing with \textit{Generative Adversarial Networks} (GANs) and to facilitate reproducibility, we have made the source code\footnote{https://github.com/fangyiyu/GNPassGAN} available to the public.
\end{itemize}

The rest of the paper is organized as follows: Section \ref{BR} reviews three categories of password guessing approaches: rule-based, probability-based, and deep learning-based, with a focus on the deep learning approaches. Section \ref{M} discusses our methodology, including the foundations of GAN, and our model GNPassGAN, while Section \ref{ER} shows our experimental design and findings. Section \ref{D} highlights limitations and future work directions.

\section{Background and Related Work}\label{BR}
This section discusses the three types of password guessing techniques as a prelude to our work.


\subsection{Rule-based Models}
The large amount of stolen passwords simplifies the process of collecting password patterns. Following that, other candidate passwords can be produced using these password patterns as guidelines. Hashcat\footnote{https://hashcat.net/wiki/} and John the Ripper\footnote{https://www.openwall.com/john/} are two popular open-source password guessing programs that provide a variety of ways for cracking passwords, including dictionary attacks, brute-force attacks, and rule-based attacks. Among all these types, the rule-based one is the fastest, and HashCat is the market leader in terms of speed, hash function compatibility, updates, and community support \cite{hranicky2019distributed}. However, rule-based systems create passwords solely based on pre-existing rules, and developing new rules requires domain expertise. Passwords that cannot be generating by applying the existing rules on existing breached passwords will hardly be guessed.

\subsection{Probability-based Models}
Two notable approaches in the probability-based password guessing category are \textit{Markov Models} and \textit{Probabilistic Context-Free Grammar} (PCFG). Markov Models are built on the assumption that all critical password features can be specified in n-grams. Its central principle is to predict the next character based on the preceding character \cite{narayanan2005fast}. PCFG examines the grammatical structures (combinations of special characters, digits, and alphanumerical sequences) in disclosed passwords and generates the distribution probability, after which it uses the distribution probability to produce password candidates \cite{weir2009password}. PCFG is actively used along with other techniques for hybrid password guessing for both targeted and trawling attacks \cite{wang2016targeted, nam2020generating, zhang2018password, liu2018genpass}. 

\subsection{Deep Learning-based Models}
Unlike rule-based or probability-based password guessing tools, deep learning-based methods make no assumptions about the password structure. The generated samples are not constrained to a particular subset of the password space. Rather than that, neural networks can autonomously encode a broad range of password information beyond the capabilities of human-generated rules and Markovian password-generating methods.  

The deep learning-based password guessing algorithms currently being explored in academia are mainly constructed of \textit{recurrent neural networks} (RNNs), autoencoders, attention mechanisms and GANs.

\subsubsection{Recurrent Neural Networks}

RNNs are neural networks in which inputs are processed sequentially and restored using internal memory. They are often employed to solve sequential tasks such as natural language processing, voice recognition, and image recognition. However, the vanilla RNN architecture is incapable of processing long-term dependencies due to vanishing gradients \cite{pascanu2013difficulty}. Thus, \textit{Long Short Term Memory} (LSTM) was designed to tackle the problem. The LSTM makes use of a gating mechanism to retain information in memory for long periods of time \cite{hochreiter1997long}.

To the best of our knowledge, Melicher et al. \cite{melicher2016fast} were the first to utilize RNNs to extract and predict password features. They kept their model, named \textit{Fast, Lean, Accurate} (FLA), as lightweight as possible in order to integrate it into local browsers for proactive password verification. Three LSTM layers and two fully connected layers comprise their proposed neural network. Various strategies for training neural networks on passwords were used. It was proven that employing transfer learning \cite{yosinski2014transferable} significantly improves guessing efficacy; however, adding natural language dictionaries to the training set and tutoring had little impact. Consequently, they discovered that FLA is superior at guessing passwords when the number of guesses is increased and when more complicated or longer password policies are targeted compared with rule-based and probability-based methods. Nevertheless, because of the Markovian nature of FLA's password generation process, any password feature that is not included within the scope of an n-gram may be omitted from encoding \cite{hitaj2019passgan}.

Zhang et al. \cite{zhang2018password} presented \textit{Structure Partition and BiLSTM Recurrent Neural
Network} (SPRNN), a hybrid password attack technique based on structural partitioning and BiLSTM. They used PCFG for structure partitioning, which seeks to structure the password training set to learn users' habit of password construction and generate a collection of basic structures and string dictionaries ordered by likelihood. The BiLSTM was then trained using the string dictionary produced by PCFG. They compared SPRNN's performance to probability-based approaches (Markov models and PCFG) on both cross-site and one-site scenarios. SPRNN outperformed the other two models in all circumstances, albeit performing worse for cross-site than for one-site.

Liu et al. \cite{liu2018genpass} developed a hybrid model named GENPass that can be generalized to cross-site attacks based on Zhang et al.'s \cite{zhang2018password} work. PCFG was also used for grammatical structure partitioning, while LSTM was used to create passwords. Additionally, they built a \textit{convolutional neural networks} (CNNs) classifier to determine which wordlist the password is most likely to originate from. The results indicate that GENPass can achieve the same degree of security as the LSTM model alone in a one-site test while generating passwords with a substantially lower rank. GENPass enhanced the matching rate by 16\% - 30\% when compared to LSTM alone in the cross-site test.

\subsubsection{Autoencoders}

Autoencoders are composed of two submodules: an encoder and a decoder. The encoder is responsible for learning the representation of the source text at each time step and generating a latent representation of the whole source sentence, which the decoder uses as an input to build a meaningful output of the original phrase. Simliar to RNNs, autoencoders are typically employed to deal with sequential data and various natural language processing tasks. RNNs and CNNs are often used as encoder and decoder components.

Pasquini et al. \cite{pasquini2021improving} applied this strategy on datasets containing leaked passwords, using GANs and \textit{Wasserstein Autoencoders} (WAEs) to develop a suitable representation of the observed password distribution rather than directly predicting it. Their methodology, called \textit{Dynamic
Password Guessing}, can guess passwords that are unique to the password set, and they are the first to apply completely unsupervised representation learning to the area of password guessing.

\subsubsection{Attention-based Models}

The attention mechanism in deep learning prioritizes certain tokens (words, letters, and phrases) while processing text inputs. This, intuitively, aids the model in gaining a better knowledge of the textual structure (e.g., grammar, semantic meaning, word structure) and hence improve text classification, generation and interpretability. In language models, the attention mechanism is often used in combination with RNNs and CNNs. Transformers \cite{vaswani2017attention} were built solely on attention, without convolution or recurrent layers. BERT \cite{devlin2018bert}, ELMO \cite{peters2018deep}, and GPT \cite{radford2019language} are all well-known instances of attention-based applications built on top of Transformers.

Li et al. \cite{li2019password} proposed a curated deep neural network architecture that consisted of five LSTM layers and an output layer. They then tutored and improved the created model using BERT. They proved that the tutoring process by BERT can help increase the model performance significantly.

\section{Methodology}\label{M}

Our model is based on the GANs model architecture and attempts to handle the most challenging difficulties in GANs in order to provide high-quality outcomes. In this section, we introduce GANs and highlight their limitations and discuss past work that used GANs for password guessing. After that, we outline our strategy and experimental design.

\begin{table*}[ht]
  \begin{center}
    \caption{Prior deep learning models used for password guessing in the literature}
    \label{tab:models}
    \begin{tabular}{llll} 
      \hline
      \textbf{Category} & \textbf{Methods} & \textbf{Models used}& \textbf{Year}\\\hline
      Autoencoders & DPG \cite{pasquini2021improving} & WAE, GAN & 2021 \\
      GAN & AdaMs \cite{272236} & GAN, HashCat & 2021 \\
      GAN & REDPACK \cite{nam2020generating} & IWGAN, RaGAN, HashCat, PCFG & 2020 \\
      GAN & PassGAN \cite{hitaj2019passgan}  & IWGAN, & 2019 \\
      Attention & Language Models \cite{li2019password}  & BERT, LSTM & 2019 \\
      RNN & GENPass \cite{liu2018genpass} & PCFG, LSTM, CNN & 2018 \\
      RNN & SPRNN \cite{zhang2018password} & PCFG, BiLSTM & 2018 \\
      RNN & FLA \cite{melicher2016fast} & LSTM & 2016 \\
      \hline
    \end{tabular}
  \end{center}
\end{table*}

\subsection{Generative Adversarial Networks}\label{GANs}

Unlike the previously described deep learning-based algorithms commonly employed in natural language processing tasks, GANs\cite{goodfellow2014generative} have been used to construct simulations of pictures, texts, and voice across all domains. Behind the scenes, a GAN consists of two sub-modules: a discriminator (D) and a generator (G), both of which are built of deep learning neural networks. G accepts noise or random features as input; learns the probability of the input's features, and generates fake data that follow the distribution of the input data. While D makes every effort to discriminate between actual samples and those created artificially by G by estimating the conditional probability of an example being false (or real) given a set of inputs (or features). This cat-and-mouse game compels D to extract necessary information from the training data. This information assists G in precisely replicating the original data distribution. D and G compete against one another during the training phase, which progressively improves their performance with each iteration. Typically, proper gradient descent and regularization techniques must be used to accelerate the whole process.

Yet, the vanilla GAN stays unstable throughout the training process because the discriminator's steep gradient space, which results in mode collapse during the generator's training phase \cite{thanh2019improving}. As a consequence, the generator is prone to deceive the discriminator before mastering the art of creating more realistic passwords. To address this,  Arjovsky et al.  \cite{arjovsky2017wasserstein} proposed \textit{Wasserstein GAN} (WGAN) which uses the earth mover distance rather than the more often employed probability distances and divergences for learning distributions. Training WGAN does not need a fine balance of discriminator and generator training, nor does it necessitate a thorough design of the network architecture. Additionally, it effectively addresses the GAN's mode collapse problem.
However, one of WGAN's requirements is that the discriminator needs to be 1-Lipschitz continuous, which indicates that the gradient's norm should never exceed 1. Arjovsky et al. \cite{arjovsky2017wasserstein} employed weight clipping in WGAN to impose a Lipschitz constraint; however, weight clipping may provide sub-optimal samples or cause the algorithm to fail to converge. Gulrajani et al. \cite{gulrajani2017improved} introduced \textit{Improved Wasserstein GAN} (IWGAN), with a new technique called gradient penalty, which punishes the discriminator's gradient norm according to its input, and enables robust training of a broad range of GAN architectures with little or no hyperparameter adjustment. 

Hitaj et al. \cite{hitaj2019passgan} proposed PassGAN, the first to implement IWGAN on password guessing. They trained the discriminator using a collection of leaked passwords (actual samples). During the training process, each iteration brings the generator's output closer to the distribution of genuine passwords, increasing the likelihood of matching real-world users' passwords. Consequently, PassGAN outperformed current rule-based password guessing tools and state-of-the-art machine learning password guessing technology FLA \cite{melicher2016fast} after sufficient passwords were generated ($10^9$).

Following the publication of PassGAN in 2019, other researchers saw the possibilities of using GANs for password guessing, and more refinements have been done on top of PassGAN. In 2020, Nam et al. \cite{nam2020generating} developed REDPACK. They modified the building block of PassGAN from ResNets to RNNs, and introduced a new cost function to stabilize the training process. In addition, they introduced a selection phase during which the password candidates are generated using several password generators. The discriminator then determines the chance of each generator's password candidates being realistic and sends the candidates with the greatest probability to password cracking tools such as HashCat. Pasquini et al. \cite{272236} proposed the \textit{Adaptive Mangling Rules Attack} (AdaMs) that applied GANs with HashCat's mangling rules to dynamically guess passwords.

We regard PassGAN to be a good representation of GANs-based password guessing tools. Following the publication of PassGAN, the majority of proposed GANs-based models in password guessing used a hybrid technique that included a rule-based model and a GANs-based model to achieve higher guessing accuracy. In reality, we can get a greater password matching accuracy by solving the intrinsic problem of IWGAN that used in PassGAN. Built on top of that, employing a hybrid model can further maximize the tool's guessing potential. 

Prior works using deep learning-based password guessing tools are sampled in Table \ref{tab:models} in chronological order. 

\begin{figure*}[t]
\centering
  \includegraphics[width=0.8\textwidth]{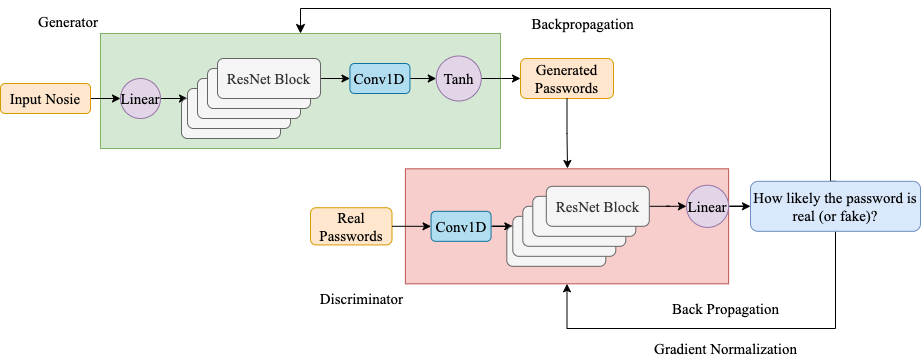}
  \caption{GNPassGAN model architecture diagram.}
  \label{fig:model}
\end{figure*} 

\subsection{Our Approach - GNPassGAN}

As mentioned in Section \ref{GANs}, the IWGAN used in PassGAN implements gradient penalty to impose the 1-Lipschitz continuous of the discriminator; however, Wu et al.\cite{GNGAN_2021_ICCV} proved that to achieve a balance between the Lipschitz restriction required by earth mover distance and the neural network’s capacity is a difficult challenge; the combination of gradient penalty and earth mover distance is not the ideal solution to GAN's mode collapse and vanishing gradient issues. To address these issues, our approach adopted a different kind of normalization technique: gradient normalization, as introduced in \cite{GNGAN_2021_ICCV}. Gradient normalization imposes a gradient norm restriction on the GANs discriminator to increase its capacity. Wu et al. \cite{GNGAN_2021_ICCV} proved that GANs trained with gradient normalization outperform previous GANs in the computer
vision area by conducting comprehensive experiments. In our study, we expect that in the password guessing domain, we can also outperform  previous methods when generating passwords with the same rank by applying gradient normalization to PassGAN.

Fig. \ref{fig:model} illustrates the architecture of our model, termed as GNPassGAN. The key improvements we made to PassGAN are as follows: We add gradient normalization \cite{GNGAN_2021_ICCV} to the discriminator, and the generator's activation function in the last layer is modified from softmax to tanh. Additionaly, instead of the Wasserstein loss, we use the binary entropy loss inside a sigmoid layer as the loss function.

\section{Evaluation}\label{ER}

Our model GNPassGAN is implemented in Pytorch 1.10. Our experiments were conducted on a workstation running Ubuntu 20.04.0 LTS, with 30 GB of RAM, an Intel(R) Xeon(R) Silver 4114 CPU, and an NVIDIA Tesla P100 GPU with 16 GB Global Memory. The hyperparameter settings for running GNPassGAN can be found in Table \ref{tab:parameters}.

\subsection{Experimental Design}
We only compare our model with PassGAN in this paper since PassGAN has conducted extensive experiments and shown that their work exceeds traditional rule-based and probability-based password guessing tools. As with PassGAN’s work, we use the \textit{rockyou} dataset\footnote{http://downloads.skullsecurity.org/passwords/rockyou.txt.bz2} for training, and the \textit{phpbb} dataset\footnote{https://github.com/danielmiessler/SecLists/blob/master/Passwords/Leaked-Databases/phpbb.txt} and a disjoint subset of \textit{rockyou} for testing. The distribution of the two datasets based on length can be found in Table \ref{tab:data_distribution}. Testing on two different sets with varying data distributions allows us to assess if our model generalizes well to cross-site password guessing. We conducted experiments on passwords of two lengths: less than or equal to 10 characters, as most password guessing experiments do; between 8 and 12 characters inclusively, which corresponds to the real-world password setting scenario as most websites require passwords to be at least 8 characters. In the following content, we refer to these two experimental settings as \textit{Char10} and \textit{Char812}, respectively. To better assess the models' guessing capability, we delete duplicates in the datasets. The training and testing sets are randomly divided by a ratio of 4:1, and there is no overlap between the two sets. Hitaj et al. \cite{hitaj2019passgan} tested their PassGAN model on passwords with less than 10 characters only; by testing on \textit{Char812}, we can see if the models are capable of properly guessing more complicated passwords. Both PassGAN and GNPassGAN are trained for 200,000 iterations in our experiment, with checkpoints for D and G retained every 10,000 iterations for the purpose of storing the neural network parameters.

\begin{table}[ht]
\centering
\caption{The comparison of the data distribution based on length in the \textit{rockyou} and \textit{phpbb} dataset. Around 95\% of passwords in \textit{phpbb} are less than or equal to 10 characters.}
\label{tab:data_distribution}
\begin{tabular}{ccc}
\hline
\textbf{Range} & \textbf{RockYou} & \textbf{phpBB} \\ \hline
(0,8)          & 33.025\%         & 48.039\%       \\ 
{[}8,10{]}     & 50.004\%         & 46.864\%       \\ 
(10,12)        & 9.906\%          & 4.031\%        \\
(12,$\infty$)     & 7.065\%          & 1.066\%        \\ 
Total          & 100\%            & 100\%          \\ \hline
\end{tabular}%
\end{table} 

Notably, the justification for using GANs to guess passwords is based on the assumption that the training and testing sets have a similar distribution, and therefore, by simulating samples from the training set, the generated samples may approximate the test set sufficiently. By shuffling before splitting to the two sets, we assume that they have a similar distribution. In Section \ref{Out}, we demonstrate empirically that our hypothesis is correct.

\begin{figure*}[ht]
\centering
\subfloat(a){\includegraphics[width=3.2in]{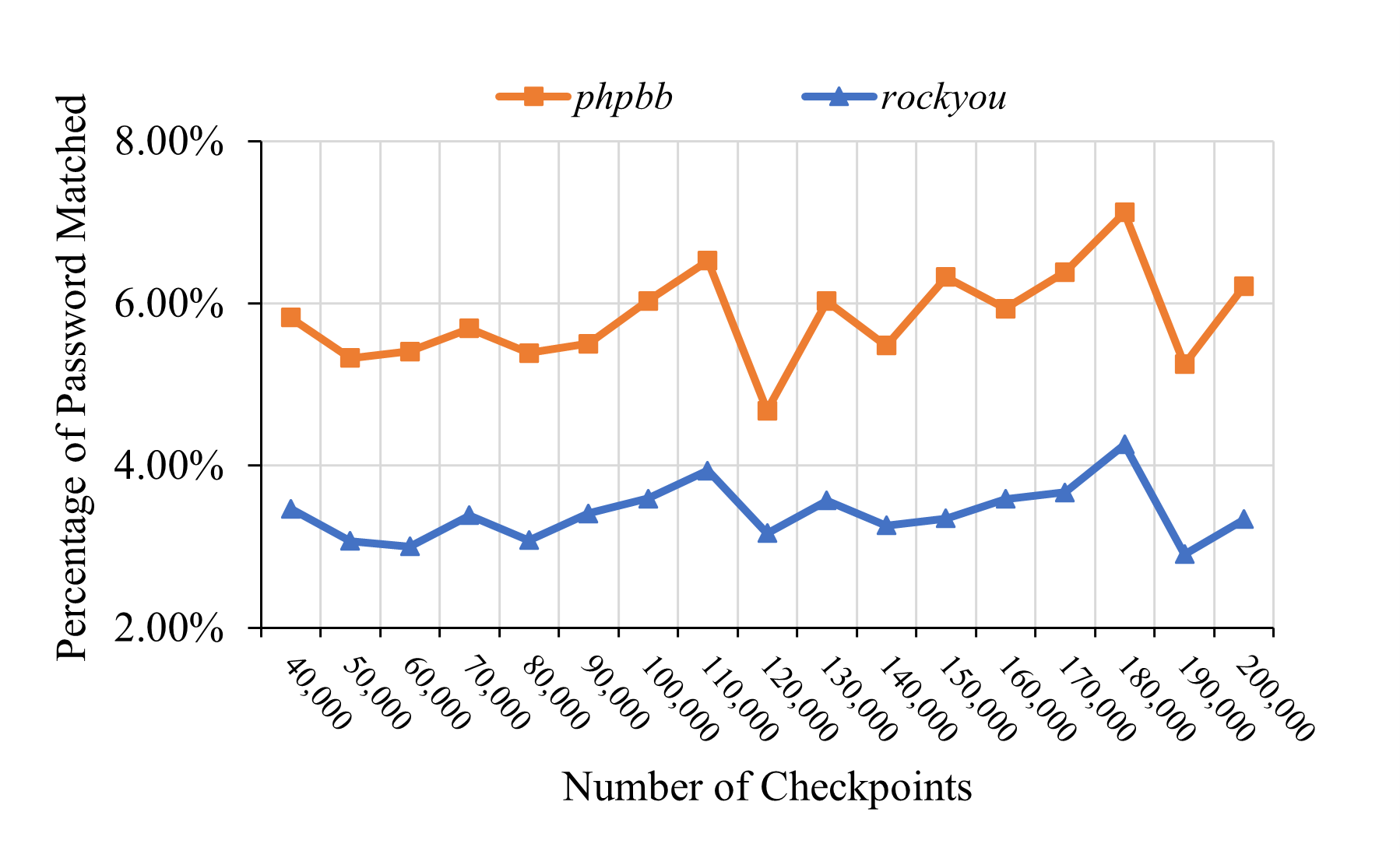}}
\hfil
\subfloat(b){\includegraphics[width=3.2in]{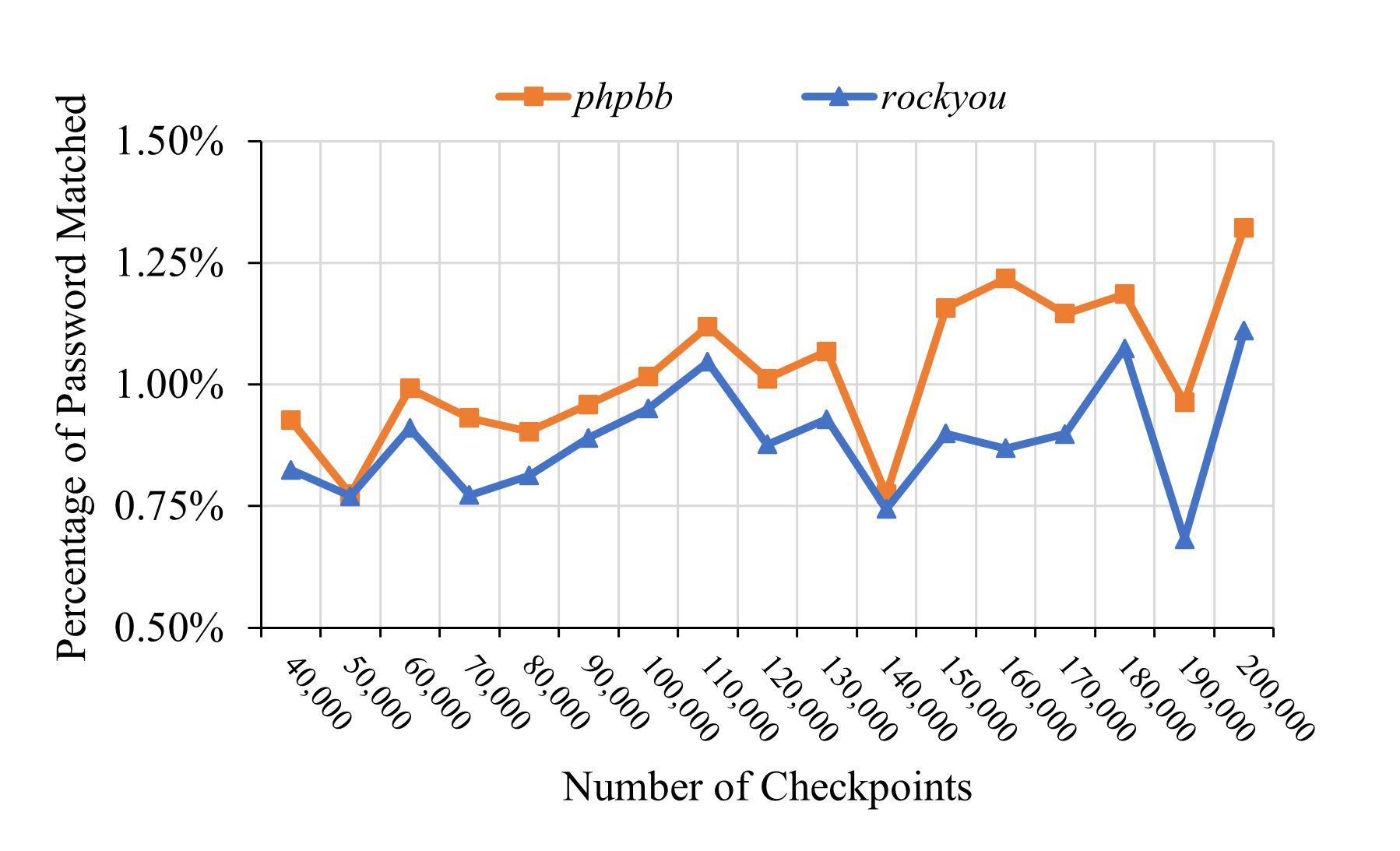}}
\caption{Proportion of unique passwords created by GNPassGAN that matched the \textit{rockyou} testing set and the \textit{phpbb} testing set at different checkpoints when experimented on passwords with length $\leqq 10$ characters (a) and in [8, 12] (b). The \textit{x} axis denotes the checkpoint used in the generating process. For each checkpoint, we sampled $10^7$ passwords.}
\label{fig:test}
\end{figure*}

\subsection{Experimental Results}

\textbf{Measuring Guessing Accuracy.}
GANs are typically applied in the computer vision area. The Inception Score \cite{salimans2016improved} and the Frechet Inception Distance \cite{heusel2017gans} are the most frequently used metrics for evaluating GAN's performance. However, since we are measuring GANs in the context of password guessing, the metrics employed in computer vision are inappropriate for our task. Instead of that, we measure the performance of GANs using the \textit{matching accuracy} as most previous work did. This metric indicates the percentage of actual passwords generated by the algorithms on an unseen dataset (test set).

Assume that the generated file is \textit{FG} and the testing file is \textit{FT}. We define the matching accuracy by dividing the number of unique passwords that exist in both \textit{FG} and \textit{FT}, by the total number of unique passwords in \textit{FT}. The following formula can be used to represent the calculation:

\begin{displaymath}
Matching\,Accuracy = \dfrac{Count(set(FG)\cap set(FT))}{Count(set(FT))}
\end{displaymath}


\textbf{GNPassGAN's Output Space.}\label{Out}
Table \ref{tab:distribution} shows the data distribution of the training set, the \textit{rockyou} testing set, and passwords produced by PassGAN and GNPassGAN on \textit{Char10} by length. As expected, the distributions of the training and testing sets are quite comparable, and the output of PassGAN and GNPassGAN both have a similar distribution to the testing set. The majority of passwords have between 5 to 10 characters. Note that a small percentage of GANs-generated passwords exceed 10 characters. This is because GANs models are attempting to simulate the distribution of the training data while also attempting to achieve sample variety.
\begin{table}[ht]
\centering
\caption{The comparison of the data distribution based on length in the training set, the \textit{rockyou} testing set and files containing fake passwords generated by PassGAN and GNPassGAN. All passwords are unique and in \textit{Char10}}
\label{tab:distribution}
\begin{tabular}{ccccc}
\hline
\textbf{Range} & \textbf{Training} & \textbf{Testing} & \textbf{PassGAN} & \textbf{GNPassGAN} \\ \hline
(0,5]    & 0.213\%  & 0.213\%  & 0.115\%  & 0.166\%  \\ 
(5,8]              & 47.661\% & 47.645\% & 46.079\% & 46.178\% \\ 
(8,10]            & 52.126\% & 52.142\% & 53.801\% & 53.645\% \\ 
(10, $\infty$)        & 0      & 0  & 0.005\%  & 0.011\%  \\ 
Total                    & 100\%    & 100\%    & 100\%    & 100\%    \\ \hline
\end{tabular}%

\end{table} 

\begin{table*}
\centering
\caption{The matched passwords by PassGAN and GNPassGAN over the \textit{rockyou} testing dataset in Char10. When $10^8$ passwords are generated, GNPassGAN is able to generate 31.69\% fewer duplicates, and match 88.03\% more passwords than PassGAN}
\label{tab:GNPassGAN vs PassGAN10}
\resizebox{\textwidth}{!}{%
\begin{tabular}{c|cc|cc}
\hline
\multirow{3}{*}{\textbf{Passwords Generated}} & \multicolumn{4}{c}{\textbf{Models}}                                  \\ \cline{2-5} 
                                    & \multicolumn{2}{c|}{\textbf{PassGAN}} & \multicolumn{2}{c}{\textbf{GNPassGAN}} \\ \cline{2-5} 
 & \textbf{Unique Passwords}& \multicolumn{1}{l|}{\textbf{Matching Accuracy}} & \multicolumn{1}{l}{\textbf{Unique Passwords}} & \multicolumn{1}{l}{\textbf{Matching Accuracy}} \\ \hline
$10^4$               & 9,738           & 103 (0.005\%)   & 9,980          &153 (0.008\%)      \\ 
$10^5$              & 94,400          & 975 (0.048\%)   & 99,545         & 1,622 (0.082\% )     \\ 
$10^6$         & 855,972         & 7,543 (0.381\%)   & 973,436        & 14,328 (0.724\%)      \\ 
$10^7$     & 7,064,483       & 40,320 (2.038\%)   & 8,806,659      & 48,263 (4.258\%)      \\ 
$10^8$      & 52,815,412      & 133,061 (6.726\%)   & 69,551,549     & 250,309 (12.647\%)     \\ \hline
\end{tabular}%
}
\end{table*}
\begin{table*}
\centering
\caption{The matched passwords by PassGAN and GNPassGAN over the \textit{rockyou} testing dataset in Char812.  When $10^8$ passwords are generated, GNPassGAN is able to generate 61.80\% fewer duplicates, and match six times more passwords than PassGAN}
\label{tab:GNPassGAN vs PassGAN812}
\resizebox{\textwidth}{!}{%
\begin{tabular}{c|cc|cc}
\hline
\multirow{3}{*}{\textbf{Passwords Generated}} & \multicolumn{4}{c}{\textbf{Models}}                                  \\ \cline{2-5} 
                                    & \multicolumn{2}{c|}{\textbf{PassGAN}} & \multicolumn{2}{c}{\textbf{GNPassGAN}} \\ \cline{2-5} 
 & \textbf{Unique Passwords}& \multicolumn{1}{l|}{\textbf{Matching Accuracy}} & \multicolumn{1}{l}{\textbf{Unique Passwords}} & \multicolumn{1}{l}{\textbf{Matching Accuracy}} \\ \hline
$10^4$               & 9,969           & 7 (0.0003\%)   & 9,983          & 26 (0.0012\%)      \\ 
$10^5$                & 98,705       & 116 (0.0055\%)   & 99,917         & 281 (0.0133\%)      \\ 
$10^6$               & 992,774         & 974 (0.0462\%)   & 995,713        & 2803 (0.1329\%)      \\ 
$10^7$         & 7,720,173      & 4,962 (0.2353\%)   & 9,672,555      & 23,416 (1.1102\%)      \\
$10^8$        & 53,025,885      & 16,404 (0.7777\%)   & 85,793,575     & 115,851 (5.4927\%)     \\ \hline
\end{tabular}%
}
\end{table*}
Both PassGAN and GNPassGAN were trained for 200,000 iterations, with the discriminator and generator competing and improving throughout each iteration. We want to see how GNPassGAN performs throughout iterations and evaluate if 200,000 is the optimal iteration parameter value for password file generation. To determine the association between iteration and matching accuracy, we display the proportion of unique passwords created by GNPassGAN that match the \textit{rockyou} and \textit{phpbb} testing sets at different checkpoints (Fig. \ref{fig:test}). As shown in Fig. \ref{fig:test} (a), the greatest matching accuracy occurs at the 180,000th checkpoints for both testing sets in \textit{Char10}. Additionally, despite the fact that our model was trained on the \textit{rockyou} dataset, the matching accuracy on the \textit{phpbb} dataset is greater than on \textit{rockyou}. A possible explanation is that the passwords in \textit{phpbb} are weaker in strength, and thus easier to guess. It also indicates that our model can perform well in cross-site guessing situations.

\textbf{Comparison with PassGAN.}
For \textit{Char10}, we generated passwords using the 180,000th checkpoints and compared them to the \textit{rockyou} testing set to determine the models' guessing capability. $10^4$ up to $10^8$ passwords were generated. Table \ref{tab:GNPassGAN vs PassGAN10} compares the matching accuracy of PassGAN with GNPassGAN, and the numbers of PassGAN are taken from the PassGAN publication \cite{hitaj2019passgan}. Hitaj et al. \cite{hitaj2019passgan} used the 200,000th iteration because PassGAN obtains the highest matching accuracy at the 200,000th iteration. It is more appropriate to compare our model's performance with theirs best performance to demonstrate which one is superior at guessing capability. It shows that as the number of created passwords increases, both models can successfully guess more passwords appearing in the test dataset. Our model GNPassGAN is capable of guessing more passwords than PassGAN and generates less duplicates, which suggests that PassGAN is experiencing mode collapsing. More precisely, when $10^8$ passwords are generated, GNPassGAN is able to match 88.03\% more passwords than PassGAN, and generate 31.69\% fewer duplicates\footnote{Both implementations took about the same time to finish the training process.}.

For \textit{Char812}, as shown in Fig. \ref{fig:test} (b), the 200,000th checkpoint has the maximum matching accuracy for both testing sets; hence, we utilize the 200,000th checkpoint to generate passwords in \textit{Char812}. The performance of PassGAN and GNPassGAN in matching passwords in \textit{Char812} is shown in Table \ref{tab:GNPassGAN vs PassGAN812}. As can be observed, GNPassGAN continues to outperform PassGAN in terms of properly guessing more genuine passwords. When $10^8$ passwords are generated, GNPassGAN is able to match six times more passwords than PassGAN, and generate 61.80\% fewer duplicates. However, when compared with Table \ref{tab:GNPassGAN vs PassGAN10}, we can find that when the same rank of passwords ($10^8$) are generated, the matching accuracy for passwords in \textit{Char812} (5.4927\%) is significantly lower than for passwords in \textit{Char10} (12.647\%), which demonstrates that lengthier passwords are more difficult to guess, and is consistent with other studies \cite{kelley2012guess} --\cite{dell2010password}. Therefore, we emphasize the importance of imposing the minimum password length restriction of eight characters to prevent passwords from being readily guessed.

\textbf{Examining Non-matched Passwords.}
We examined a list of GNPassGAN-generated passwords that did not match any of the testing sets and discovered that a substantial number of these passwords are plausible candidates for human-generated passwords. As a result, we expect that the passwords created by GNPassGAN might be exploited as honeyword candidates to reduce attackers' success rate at compromising users' accounts and detect data breaches. 

Honeywords were introduced by Juels and Rivest as a potential method for efficiently detecting password leaks \cite{juels2013honeywords}. According to their proposal, a website could store decoy passwords, called honeywords, alongside real passwords in its credential database, so that even if an attacker steals and reverts the password file containing the users’ hashed passwords, they must still choose a real password from a set of distinct sweetwords (a real password and its associated honeywords are referred to as sweetwords). The attacker’s use of a honeyword could cause the website to become aware of the breach. Notably, honeywords are only beneficial if they are difficult to distinguish from real-world passwords; otherwise, a knowledgeable attacker may be able to recognize them and compromise their security. Table \ref{tab:unmatched} illustrates some samples of the passwords
generated by GNPassGAN that did not match the testing set but seem to be viable honeyword candidates.
\begin{table}[ht]
\centering
\caption{Passwords produced by GNPassGAN that did not match the testing sets.}
\label{tab:unmatched}
\begin{tabular}{cccc}
\hline
claia02001   & cas043712 & mannda235 & all53002  \\
badanan24 & nsha1105  & livemilo  & namrasbdo \\
mintesa01 & jonern14  & tikiocmo  & dendiona  \\
maketa11  & moritin1  & pilk2711  & fish1053   \\
\hline
\end{tabular}%
\end{table}

\section{Discussion}\label{D}

\subsection{Limitation and Future Work}
This section discusses the limitations of GNPassGAN, the future work that can be done to enhance GNPassGAN, and potential applications.

\textbf{Inappropriate password setting.} PassGAN and GNPassGAN both conduct experiments with passwords that are less than or equal to 10 characters in length, with nearly half of them being less than or equal to 8 characters. This is unworkable in practice, since the majority of websites need a minimum of 8 characters. Additionally, some websites require a mix of numeric characters, special symbols, and special characters in passwords. Previously published research \cite{tan2020practical} shows that altering password criteria, such as required minimum length and class, may have a considerable positive influence on both usability and security. As a result, while preprocessing datasets and assessing the models, we need to take the password policy and minimum length requirements into account to simulate real-world password generating scenarios.

\textbf{Hybrid models.} Given that prior work \cite{272236} and \cite{hitaj2019passgan} demonstrated that deep neural networks can mimic the domain knowledge of professional attackers, GNPassGAN can be used in conjunction with rule-based models such as HashCat to provide more accurate dynamic password guessing solutions than GNPassGAN alone.

\textbf{Honeyword generation.} Because GNPassGAN is capable of synthesizing texts with the same distribution as the training data, it can generate authentic-looking passwords that can be considered as honeyword candidates when trained on real-world password datasets. We are now experimenting with the use of GNPassGAN to generate honeywords and doing quantitative evaluations on the honeywords' quality in terms of attack success rate.

\subsection{Conclusion}
In this paper, we compared the commonly deployed rule-based, probability-based password guessing tools with deep learning-based approaches. We particularly discussed the deep learning approaches utilized, including RNNs, autoencoders, attention mechanisms, and GANs. Additionally, we introduced GNPassGAN, a GANs-based deep learning password guessing tool. The original motivation comes from PassGAN \cite{hitaj2019passgan}, and by applying gradient normalization to the discriminator, modifying the loss function, and tweaking the architecture of the generator, we are able to outperform PassGAN by 88.03\% while generating $10^8$ passwords and create 31.69\% less duplicates. The result indicates that GNPassGAN is superior than PassGAN in terms of resolving the mode collapse issue and achieving a better guessing capability. We argue that there is no need to compare GNPassGAN to traditional rule-based and probability-based password guessing tools, given that Hitaj et al. \cite{hitaj2019passgan}, the authors of PassGAN have conducted extensive experiments and shown that their work outperforms others. 

We encourage researchers interested in password guessing with deep learning techniques to adopt GNPassGAN as a new state-of-the-art benchmark. Additionally, the potential for using GNPassGAN to construct password strength meters that encourage users to create stronger passwords, and honeywords that detect password breaches is promising.

\section{Acknowledgement}
The authors acknowledge the support of the Natural Sciences and Engineering Research Council of Canada (NSERC), funding reference number RGPIN-2018-05919.

\bibliographystyle{IEEEtran}
\bibliography{References}
\begin{appendix}
\section{Hyperparameters for Running GNPassGAN}
Our GNPassGAN was trained using the following hyperparameter settings.
\begin{table}[h]
\centering
\caption{Hyperparameter Setting for Trainning GNPassGAN}
\label{tab:parameters}
\begin{tabular}{ccc}
\hline
\textbf{Hyperparameters}                        & \textbf{Value} \\ \hline
Batch size                                      & 64             \\ 
Number of iterations                            & 200,000        \\ 
\begin{tabular}[c]{@{}c@{}}Number of discriminator iterations\\  for each generator iteration\end{tabular} & 10 \\ 
Layer dimension for generator and discriminator & 128            \\ 
Adam learning rate                              & 0.0001         \\ 
Adam coefficient $\beta_1$                          & 0.5            \\ 
Adam coefficient $\beta_2$                    & 0.9            \\ \hline
\end{tabular}%
\end{table}
\end{appendix}
\end{document}